\documentclass[aps,prd,twocolumn,showpacs,amsmath,amssymb,nofootinbib,
superscriptaddress]{revtex4-1}

\usepackage{epsfig}
\usepackage{graphicx}
\usepackage{dcolumn}
\usepackage{latexsym}
\usepackage{graphicx}
\usepackage{enumerate}
\usepackage{float}
\usepackage{placeins}
\usepackage{supertabular}

\usepackage{hyperref} 
\hypersetup{
   colorlinks=true,
   linkcolor=blue,
   citecolor=blue,
   urlcolor=blue
}

\def\tstrut{\vrule height2.25ex depth0pt width0pt} 

\bibpunct{[}{]}{,}{n}{,}{,} 

\begin{document}

\title{$B$ decays into radially excited charmed mesons}
\author{J. Segovia}\affiliation{Physics Division, Argonne National Laboratory, \\
Argonne, Illinois 60439-4832, USA}
\author{E. Hern\'andez}\affiliation{Departamento de F\'{\i}sica Fundamental e IUFFyM,
\\ Universidad de Salamanca, E-37008 Salamanca, Spain}
\author{F. Fern\'andez}\affiliation{Departamento de F\'{\i}sica Fundamental e IUFFyM,
\\ Universidad de Salamanca, E-37008 Salamanca, Spain}
\author{D.R. Entem}\affiliation{Departamento de F\'{\i}sica Fundamental e IUFFyM, \\
Universidad de Salamanca, E-37008 Salamanca, Spain}

\date{\today}

\begin{abstract}
It has been recently argued that some longstanding problems in semileptonic $B$
decays can be solved provided the branching ratio for the $B\to D^{\prime(\ast)}$
semileptonic decays are large enough. We have studied these decays in a constituent
quark model which has been successful in describing semileptonic and non-leptonic $B$
decays into orbitally excited charmed mesons. Our results do not confirm the
hypothesis of large branching ratios for the $B\to D^{\prime(\ast)}$ semileptonic
decays. In addition, we calculate the non-leptonic $B\to D'\pi$ decays which can
provide an independent test of the form factors involved in the $B\to
D^{\prime(\ast)}$ reactions.
\end{abstract}

\pacs{12.39.Jh, 12.39.Pn, 13.20.He, 14.40.Lb}
\keywords{potential models, properties of charmed mesons, leptonic and semileptonic
decays.}

\maketitle

\section{INTRODUCTION}
\label{sec:introduction}

The determination of exclusive branching fractions of $B\to X_{c}l^{+}\nu_{l}$ decays
is an essential part of the $B$-factory program to understand the dynamics of
$b$-quark semileptonic decays and therefore to determine the relevant
Cabibbo-Kobayashi-Maskawa matrix elements, $|V_{cb}|$ and $|V_{ub}|$, which are
essential parameters of the Standard Model. The methods used to determine the 
decay rates as a function of $|V_{qb}|$ are very different for inclusive and
exclusive decays, and the comparison of these complementary determinations provides
an important check on the results.

Current measurements show a discrepancy between the inclusive rate ${\cal B}(B^{+}\to
X_{c}l^{+}\nu_{l})$ and the sum of the measured exclusive channels. At present, the
disagreement between the inclusive rate ${\cal B}(B^{+}\to X_{c}
l^{+}\nu)=(10.92\pm0.16)\%$ and the sum of the measured exclusive rates is
\begin{equation}
\begin{split}
&
{\cal B}(B^{+}\to X_{c}l^{+}\nu_{l}) - {\cal B}(B^{+}\to D^{(\ast)}l^{+}\nu_{l}) \\
&
-{\cal B}(B^{+}\to D^{(\ast)}\pi l^{+}\nu_{l}) = (1.45\pm0.13)\%.
\label{eq:discrepancy}
\end{split}
\end{equation}

The authors of Ref.~\cite{Bernlochner:2012bc} argue that this $1.45\%$ discrepancy
could be solved, or at least eased, by an unexpectedly large $B$ decay rate to the
first radially excited $D^{\prime}$ and $D^{\prime\ast}$ states (${\cal B}(B\to
D^{\prime(\ast)}l^{+}\nu_{l})\sim{\cal O}(1\%)$). It is also stated in 
Ref.~\cite{Bernlochner:2012bc} that a potentially large ${\cal B}(B^{+}\to
D^{\prime(\ast)} l^{+}\nu_{l})$ could help to solve the so called $1/2$ vs $3/2$
puzzle. This puzzle refers to the discrepancy between heavy quark symmetry based
model predictions for $B$ decays into $1P$ $D^{\ast\ast}$ states and observations.
Those calculations predict that these decays should have a substantially smaller rate
to the $j_{q}^{P}=\frac{1}{2}^{+}$ doublet than to the $j_{q}^{P}=\frac{3}{2}^{+}$
doublet (see for instance Refs.~\cite{PhysRevD.56.5668,Bigi2007}) whereas roughly
equal branching ratios are found experimentally. A large  ${\cal B}(B^{+}\to
D^{\prime(\ast)} l^{+}\nu_{l})$ would result in an excess of the detected $B\to
D_{1/2}\pi l \nu_{l}$ with respect to $B\to D_{3/2}\pi l \nu$ due to the fact that
the $\Gamma(D^{\prime}\to D_{1/2}\pi)$ is much larger than the $\Gamma(D^{\prime}\to
D_{3/2}\pi)$ because in the first case the pion is emitted in $S$-wave while in the
$D_{3/2}$ one is emitted in $P$-wave.

Evidence for the $D^{\prime(\ast)}$ radial excitations, that would correspond to $2S$
quark model states, have recently been found by the BaBar
Collaboration~\cite{delAmoSanchez:2010vq}. Their results are summarized in
Table~\ref{tab:experimentalDast}. The $D(2550)$ meson has only been seen in the decay
mode $D^{\ast}\pi$ and its helicity-angle distribution turns out to be consistent
with the prediction of a $2^{1}S_{0}$ state. The $D^{\ast}(2600)$ meson decays into
$D\pi$ and $D^{\ast}\pi$ final states, and its helicity-angle distribution is
consistent with the meson being a $J^{P}=1^{-}$ state. Moreover, its mass makes it
the perfect candidate to be the spin partner of the $D(2550)$ meson.

\begin{table}[!t]
\begin{center}
\begin{tabular}{lcc}
\hline
\hline
Resonance & Mass (MeV) & Width (MeV) \\
\hline
\tstrut
$D(2550)^{0}$ & $2539.4\pm4.5\pm6.8$ & $130\pm12\pm13$ \\
$D^{\ast}(2600)^{0}$ & $2608.7\pm2.4\pm2.5$ & $93\pm6\pm13$ \\
\hline
\hline
\end{tabular}
\caption{\label{tab:experimentalDast} Mass and total decay width of the
$D^{\prime(\ast)}$ resonances as measured by the BaBar
Collaboration~\cite{delAmoSanchez:2010vq}.}
\end{center}
\end{table}
 
Previous theoretical determinations of ${\cal B}(B^{+}\to D^{\prime(\ast)} l^{+}\nu)$
include an earlier calculation within the ISGW2 model, which incorporates constraints
imposed by heavy quark symmetry, where a value of $0.06\%$ was
obtained~\cite{Scora:1995ty}. This is in agreement with the heavy quark effective
theory calculation of Suzuki {\it et al.}~\cite{Suzuki:1993iq} which obtained ${\cal
B}(B^{+}\to D^{\prime(\ast)}l^{+}\nu)=0.05\%$. A much larger value is obtained in
Ref.~\cite{Ebert:1999ed}. There the authors use a relativistic quark model finding
that the calculated branching ratio increases from $0.29\%$, for infinite heavy quark
mass, to $0.40\%$ when $1/m_q$ corrections are  taken into account. Although the
value is still far from the $\sim\!\!1\%$ needed to explain the experimental
discrepancy, the size of the corrections suggests that a complete calculation may
further approach the $1\%$ result.

In this work we shall perform a full determination of ${\cal B}(B^{+}\to D^{\prime(\ast)}
l^{+}\nu)$  within the framework of the constituent quark
model described in Ref.~\cite{vijande2005constituent}. The model has recently been
applied to mesons containing heavy quarks obtaining a satisfactory description of
many physical observables: spectra~\cite{PhysRevD.78.114033,PhysRevD.80.054017},
strong decays and reactions~\cite{Segovia:2012cd,Segovia:2011zza}, and semileptonic
and nonleptonic $B$ and $B_{s}$ decays into orbitally excited charmed and
charmed-strange mesons~\cite{segovia2011semileptonic,PhysRevD.86.014010}. We will use
the factorization approximation which gave a satisfactory explanation of the decays
analyzed in Refs.~\cite{segovia2011semileptonic,PhysRevD.86.014010}. The form factors
that parametrize the $\Gamma(B\to D^{\prime}l\nu)$ decay also appear, evaluated  at
$q^{2}=m_{\pi}^{2}$, in the non-leptonic decay $B\to D^{\prime}\pi$ evaluated in
factorization approximation. The latter decay, if experimentally accessible, could be
used to extract  information on the form factors near $q^{2}=0$. In this work we
also evaluate the $\bar{B}^{0}\to D^{\prime+}\pi^{-}$ branching ratio and its
branching fraction relative to the $\bar{B}^{0}\to D^{+}\pi^{-}$ decay.

The paper is organized as follows: In Sec.~\ref{sec:CQM}, we describe the constituent
quark model predictions for the first radially excited $S$-wave states paying special
attention to their strong decays. Sec.~\ref{sec:weakdecays} is dedicated to explain
the theoretical framework through which we calculate the ${\cal B}(B^{+}\to
D^{\prime(\ast)}l^{+}\nu)$ and ${\cal B}(\bar B^{0}\to D^{\prime +}\pi^{-})$. Our results
are shown in Sec.~\ref{sec:results}. We summarize our conclusions in
Sec.~\ref{sec:conclusions}.

\section{CONSTITUENT QUARK MODEL PREDICTIONS FOR $2S$ STATES}
\label{sec:CQM}

All the details on the constituent quark model we use have been described in
Ref.~\cite{vijande2005constituent} and we will only sketch here its main features.
The model is based on the assumption that the constituent quark mass of the light
quarks is due to the spontaneous chiral symmetry breaking of the QCD Lagrangian. To
restore the original symmetry an interaction term, due to Goldstone-Boson exchanges,
appears between light quarks. This interaction is added to the perturbative One-Gluon
Exchange (OGE) and the non-perturbative confining interactions. In the heavy quark
sector, chiral symmetry is explicitly broken and Goldstone-boson exchanges do not
appear. Therefore, the corresponding potential for the system stems from the
nonrelativistic reduction of the OGE interaction and the confinement component.
Explicit formulae and model parameters used herein can be find in
Refs.~\cite{PhysRevD.78.114033,segovia2011semileptonic}.

Concerning the new discovered mesons, $D(2550)$ and $D^{\ast}(2600)$, we predict
masses of $2.70\,{\rm GeV}$ and $2.75\,{\rm GeV}$ which are larger than the
experimental ones, $2.54\,{\rm GeV}$ and $2.61\,{\rm GeV}$. While masses are large,
the goodness of the meson wave functions has been tested through their strong decays.
Those were evaluated in Ref.~\cite{Segovia:2012cd} in which a modification
of the phenomenological $^{3}P_{0}$ decay model was proposed. The strength $\gamma$ 
of the decay interaction is scale-dependent being a function of the reduced mass of
the quark-antiquark pair of the decaying meson. In this way a satisfactory global
description of strong decays of mesons which belong to charmed, charmed-strange,
hidden charm and hidden bottom sectors was achieved~\cite{Segovia:2012cd}. Using
physical masses for the $2S$ states, we obtained the results shown in
Tables~\ref{tab:D2550} and~\ref{tab:D2600}.

\begin{table}[!t]
\begin{center}
\begin{tabular}{ccc}
\hline
\hline
\multicolumn{3}{c}{$D(2550)$ as $nJ^{P}=2\,0^{-}$} \\
\hline
Channel & $\Gamma_{^{3}P_{0}}$ & ${\mathcal B}_{^3P_0}$ \\[2ex]
$D^{\ast}\pi$ & $131.90$ & $99.87$ \\
$D_{0}^{\ast}\pi$ & $0.18$ & $0.13$ \\
total & $132.07$ & $100$ \\
\hline
\hline
\end{tabular}
\caption{\label{tab:D2550} Open-flavor strong decay widths, in MeV, and branchings,
in $\%$, of the $D(2550)$ meson with quantum numbers $nJ^{P}=2\,0^{-}$.}
\end{center}
\end{table}

\begin{table}[t!]
\begin{center}
\begin{tabular}{ccc}
\hline
\hline
\multicolumn{3}{c}{$D^{\ast}(2600)$ as $nJ^{P}=2\,1^{-}$} \\
\hline
Channel & $\Gamma_{^{3}P_{0}}$ & ${\mathcal B}_{^3P_0}$ \\[2ex]
$D\pi$ & $10.84$ & $11.19$ \\
$D^{\ast}\pi$ & $54.10$ & $55.83$ \\
$D\eta$ & $11.86$ & $12.24$ \\
$D_{s}K$ & $8.73$ & $9.01$ \\
$D^{\ast}\eta$ & $9.65$ & $9.95$ \\
$D_{1}\pi$ & $0.28$ & $0.29$ \\
$D'_{1}\pi$ & $1.44$ & $1.49$ \\
$D_{2}^{\ast}\pi$ & $0.01$ & $0.00$ \\
total & $96.91$ & $100$ \\
\hline
\hline
\end{tabular}
\caption{\label{tab:D2600} Open-flavor strong decay widths, in MeV, and branchings,
in $\%$, of the $D^{\ast}(2600)$ meson with quantum numbers $nJ^{P}=2\,1^{-}$.}
\end{center}
\end{table}

The $D(2550)$ meson has been only seen in the decay mode $D^{\ast}\pi$, thus its
possible spin-parity quantum numbers up to $J=3$ are $J^{P}=0^{-}$, $1^{+}$, $2^{-}$
and $3^{+}$. It is the lower in mass of the newly discovered mesons and within the
possible assignments, the $0^{-}$ is the most plausible. Assuming it is a
$2S,J^{P}=0^{-}$ state, its total width, shown in Table~\ref{tab:D2550}, predicted
by the $^{3}P_{0}$ model is in very good agreement with the experimental value given
by the BaBar Collaboration.

The $D^{\ast}(2600)$ meson decays into $D\pi$ and $D^{\ast}\pi$ final states. Thus,
its possible quantum numbers are $J^{P}=1^{-}$, $2^{+}$ and $3^{-}$. The
helicity-angle distribution of $D^{\ast}(2600)$ is found to be consistent with
$J^{P}=1^{-}$. Moreover, its mass makes it the perfect candidate to be the spin
partner of the $D(2550)$ meson. Again, the total decay width predicted by the
$^{3}P_{0}$ model, assuming it is a $2S,J^{P}=1^{-}$ state, is in good agreement with
the experimental data. Besides, we predict the ratio of branching fractions 
\begin{equation}
\frac{{\cal B}(D^{\ast}(2600)^{0} \to D^{+}\pi^{-})}
{{\cal B}(D^{\ast}(2600)^{0} \to D^{\ast+}\pi^{-})} = 0.20
\end{equation}
in reasonable agreement with experiment,
$0.32\pm0.02\pm0.09$~\cite{delAmoSanchez:2010vq}.

In Tables~\ref{tab:D2550} and~\ref{tab:D2600} we also show the contribution of each
channel to the total decay width of the $D^{\prime(\ast)}$ states. In both cases we
obtain that the decay widths of these mesons into $1P$ states are much smaller than
the ones into $1S$ states. This is in contrast to Ref.~\cite{Bernlochner:2012bc}
where the authors find plausible that the $D^{\prime(\ast)}$ decay rates to $1S$ and
$1P$ charmed states may be comparable. That was used in
Ref.~\cite{Bernlochner:2012bc} as a possible explanation of the so called $1/2$
versus $3/2$ puzzle. However, we studied this issue within our model in
Ref.~\cite{segovia2011semileptonic} where we observed similar $B$ semileptonic decay
rates into the two $j_{q}^{P}=1/2^{+}$ and $j_{q}^{P}=3/2^{+}$ doublets, being our
results for the different channels in agreement with experiment.

\section{THE $B^{+}\to D^{\prime(\ast)}l^{+}\nu_{l}$ AND $B^{0}\to
D^{\prime(\ast)}\pi$ DECAY WIDTH}
\label{sec:weakdecays}

The presence of the heavy quark in the initial and final meson states in these decays
considerably simplifies their theoretical description. Let us start our analysis in
the infinitely heavy quark limit, $m_{Q}\to \infty$. In this limit the heavy quark
symmetry arises. This leads to a considerable reduction of the number of independent
form factors which are necessary for the description of heavy-to-heavy semileptonic
decays. For example, in this limit only one form factor is necessary for the
semileptonic $B$ decay to $S$-wave $D$ mesons. It is important to note that the heavy
quark symmetry requires that matrix elements between a $B$ meson and an excited $D$
meson should vanish at zero recoil as a result of the orthogonality of the wave
functions.
 
As the $D'$ and $D^{\prime\ast}$ states have $0^{-}$ and $1^{-}$ quantum numbers,
the hadronic matrix elements for the semileptonic transition can be parameterized in
terms of form factors as~\cite{Manohar:2000dt}
\begin{widetext}
\begin{equation}
\begin{split}
\frac{\langle D'(p')|\bar{\Psi}_{c}(0)\gamma^{\mu}(1-\gamma_{5})\Psi_{b}(0)|B(p)\rangle}
{\sqrt{m_{B}m_{D'}}} =& h_{+}(w) (v+v')^{\mu} + h_{-}(w)(v-v')^{\mu}, \\
\frac{\langle
D^{\prime\ast}(p')|\bar{\Psi}_{c}(0)\gamma^{\mu}(1-\gamma_{5})\Psi_{b}(0)|B(p)\rangle}
{\sqrt{m_Bm_{D^{\prime\ast}}}} =& h_{v}(w)\epsilon^{\mu\nu\alpha\beta}
\epsilon^{\ast}_{\nu} v'_{\alpha} v_{\beta} \\
&
- i [-h_{A_{1}}(w) (w+1)\epsilon^{\ast\mu}+ h_{A_{2}}(w)(\epsilon^{\ast}\cdot v)
v^{\mu} + h_{A_{3}}(w)(\epsilon^{\ast}\cdot v) v^{\prime\mu}],
\end{split}
\end{equation}
\end{widetext}
where $v(v')$ is the four velocity of the $B(D^{\prime(\ast)})$ meson, 
$\epsilon^{0123}=-1$, $\epsilon^{\mu}$ is a polarization vector of the final vector
charmed meson and the form factors $h_{i}$ are dimensionless functions of the product
of four-velocities $w=v\cdot v'$. The differential $d\Gamma/dw$ decay widths are
given by~\cite{Manohar:2000dt}
\begin{widetext}
\begin{equation}
\begin{split}
\frac{d\Gamma}{dw}(B^{+}\to D'l^{+}\nu_{l}) &=
\frac{G_{F}^{2} |V_{cb}|^2 m_{B}^{5}}{48\pi^{3}}(w^{2}-1)^{3/2}r^{3}(1+r)^{2}
G^{2}(w), \\
\frac{d\Gamma}{dw}(B^{+}\to D^{\prime\ast}l^{+}\nu_{l}) &=
\frac{G_{F}^{2} |V_{cb}|^2 m_{B}^{5}}{48\pi^{3}}(w^{2}-1)^{3/2}
(w+1)^{2} r^{\ast3}(1-r^{\ast})^{2} \left[1+\frac{4w}{w+1}
\frac{1-2wr^{\ast}+r^{\ast2}}{(1-r^{\ast})^{2}}\right] F^{2}(w),
\end{split}
\end{equation}
\end{widetext}
where $G_{F}$ is the Fermi decay constant, $|V_{cb}|$ the modulus of the
Cabibbo-Kobayashi-Maskawa matrix element for a $b\to c$ transition and 
$r^{(\ast)}=m_{D^{\prime(\ast)}}/m_{B}$. As for $G(w)$ and $F(w)$, they are given in
term of the form factors as
\begin{widetext}
\begin{equation}
\begin{split}
G(w) =& h_{+}(w) - \frac{1-r}{1+r} h_{-}(w), \\
F^{2}(w) =& \bigg\{2(1-2wr^{\ast} + r^{\ast2}) \left[h^{2}_{A_{1}}(w) +
\frac{w-1}{w+1} h_{V}^{2}(w) \right] \\
&
+ \bigg[(1-r^{\ast})h_{A_{1}}(w) + (w-1) \big(h_{A_{1}}(w) - h_{A_{3}}(w) -
r^{\ast} h_{A_{2}}(w)\,\big) \bigg]^{2} \bigg\} 
\times \bigg\{(1-r^{\ast})^{2} + \frac{4w}{w+1} (1 - 2wr^{\ast} +
r^{\ast2})\bigg\}^{-1}.
\end{split}
\end{equation}
\end{widetext}
Similarly to the semileptonic $B^{+}\to D^{(\ast)}l^{+}\nu_{l}$ case we have that, in
the limit of very large heavy quark masses, heavy quark symmetry predicts that all
form factors are given in terms of just one Isgur-Wise function $\xi(w)$. One has in
that limit
\begin{equation}
h_+(w)=h_V(w)=h_{A_1}(w)=h_{A_3}(w)=\xi(w),
\label{eq:xi}
\end{equation}
while
\begin{equation}
h_-(w)=h_{A_2}(w)=0.
\label{eq:0}
\end{equation}
From these results one obtains in that limit
\begin{equation}
G(w)=F(w)=\xi(w).
\end{equation}
 
Heavy quark symmetry at the point of zero recoil $(\omega=1)$ implies $\xi(1)=0$ in
the $B\to D^{\prime (\ast)}$ case since the radial parts of the wave functions for
the $D^{\prime (\ast)}$ and $B$ mesons are orthogonal in the infinitely heavy quark
mass limit. Thus, the value of $\xi(\omega)$ near zero recoil comes entirely from
corrections beyond that limit. This is different from the $B\to D^{(\ast)}$ case
where $\xi(1)=1$ in that limit.

As mentioned in Ref.~\cite{Bernlochner:2012bc}, the nonleptonic $\bar
B^{0}\to D^{\prime+}\pi^{-}$ decay could also give valuable information on $F(w)$ and
$G(w)$, as in factorization approximation the width of this process is related to the
form factors involved in the semileptonic decay. Factorization approximation has
been proven to be correct for $B\to D\pi$ in the infinite heavy quark mass
limit~\cite{PhysRevLett.87.201806}, and we expect it should also work for decays $B$
into $D^{\prime}\pi$.

Following Ref.~\cite{Becirevic:2013mp}, we calculate the ratio ${\cal B}(\bar
B^{0}\to D^{\prime+}\pi^{-})/{\cal B}(\bar B^{0}\to D^{+}\pi^{-})$ which in
factorization approximation is given by
\begin{equation}
\begin{split}
\frac{{\cal B}(\bar B^{0}\to D^{\prime+}\pi^{-})}{{\cal B}(\bar B^{0}\to
D^{+}\pi^{-})} =&
\left(\frac{m^{2}_{B}-m^{2}_{D^{\prime}}}{m^{2}_{B}-m^{2}_{D}}\right)^{2} \times \\ 
&
\hspace{-1.30cm} \times \left(\frac{\lambda\left(m_{B}^{2},m_{D^{\prime}}^{2},
m_{\pi}^{2}\right)}{\lambda\left(m^{2}_{B},m^{2}_{D},m^{2}_{\pi}\right)}\right)^{1/2}
\left| \frac{f^{B\to D^{\prime}}_{+}(0)}{f^{B\to D}_{+}(0)} \right|^{2},
\end{split}
\end{equation}
where $\lambda(a,b,c)=(a+b-c)^2-4ab$,  and the $f_{+}(q^2)$ form factor 
is related with $h_{\pm}$ via
\begin{equation}
\begin{split}
f_{+}(q^2) &=
\frac{1}{2} \left(\sqrt{\frac{m_{D^{\prime}}}{m_{B}}} +
\sqrt{\frac{m_{B}}{m_{D^{\prime}}}}\right) h_{+}(\omega) \\
&
+ \frac{1}{2} \left(\sqrt{\frac{m_{D^{\prime}}}{m_{B}}} -
\sqrt{\frac{m_{B}}{m_{D^{\prime}}}}\right) h_{-}(\omega).
\end{split}
\end{equation}
Note $q^{2}$ and $\omega$ are related through
$q^{2}=m_{B}^{2}+m_{D^{(\prime)}}^{2}-2m_{B}m_{D^{(\prime)}}\omega$. If the ${\cal
B}(\bar B^{0}\to D^{\prime+}\pi^{-})$ branching ratio is measured experimentally, it
could be used to extract information on $f_{+}^{B\to D'}(0)$.
 
\section{RESULTS}
\label{sec:results}

\begin{figure}[!t]
\resizebox{7.cm}{!}{\includegraphics{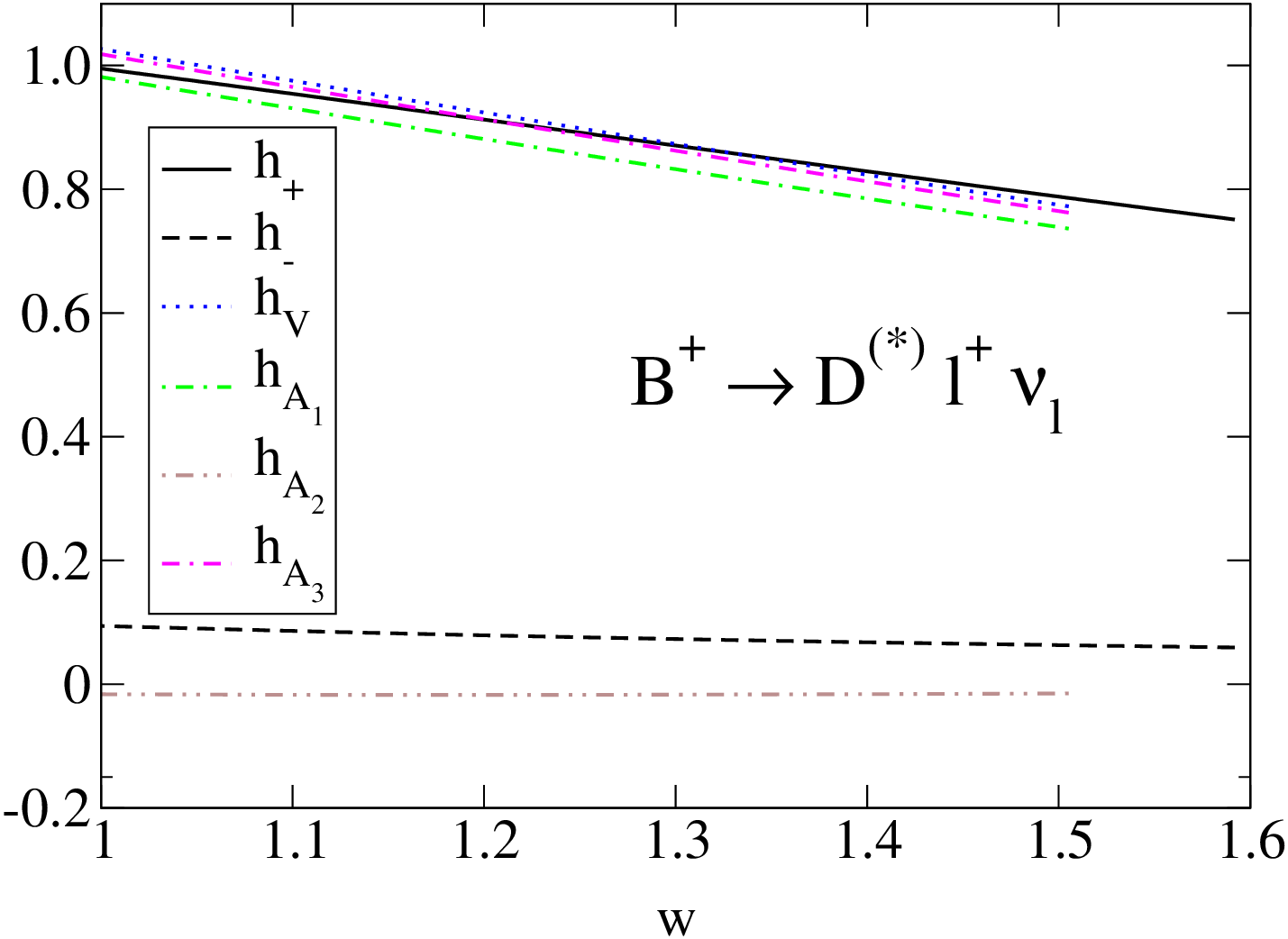}} \vspace{.8cm} \\
\resizebox{7.cm}{!}{\includegraphics{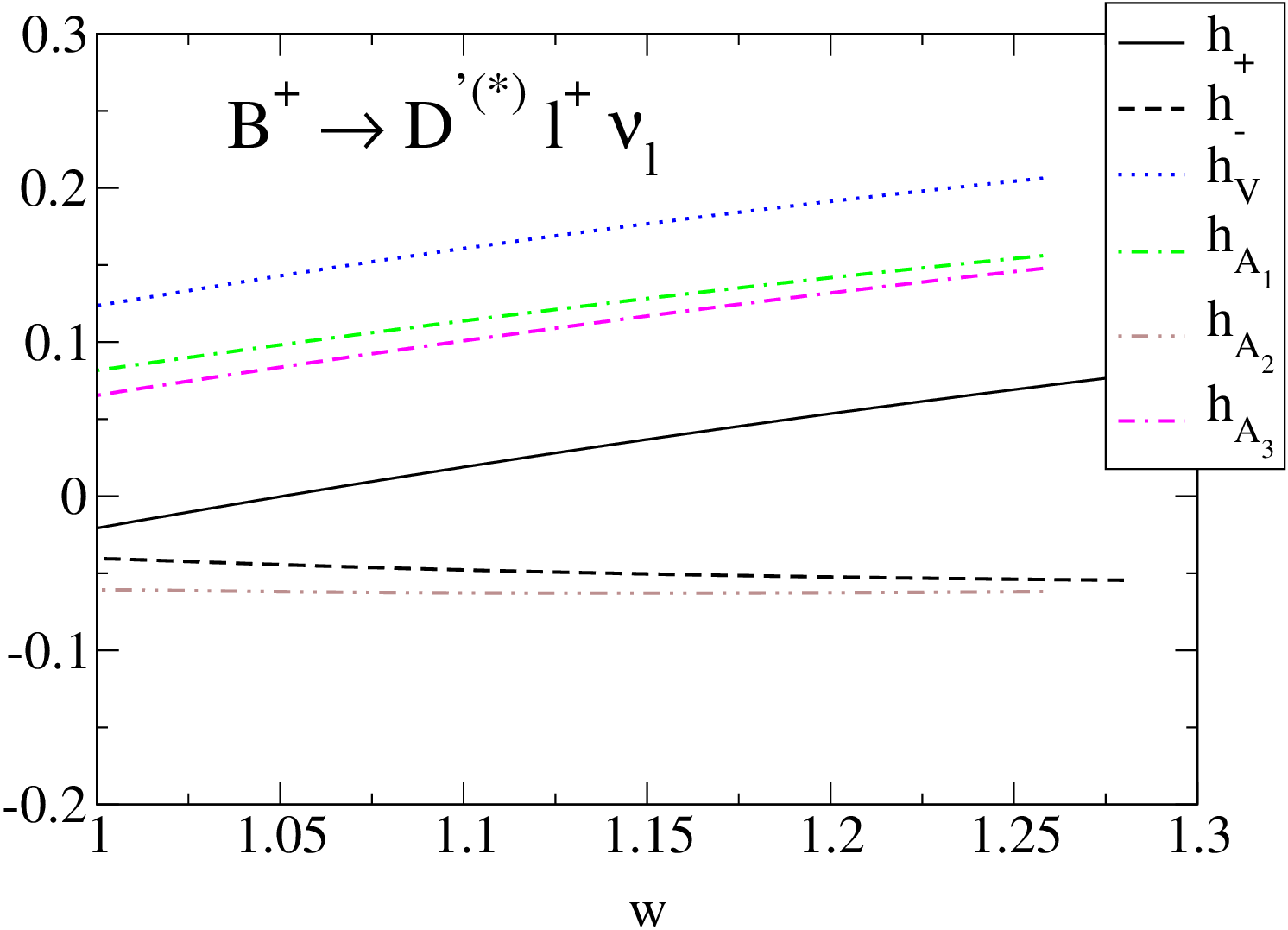}}
\caption{\label{fig:ff} Upper panel: Form factors for the $B^{+}\to
D^{(\ast)}l^{+}\nu_{l}$ transition evaluated in our model. Lower panel: The same for
the $B^{+}\to D^{\prime(\ast)}l^{+}\nu_{l}$ transition.}
\end{figure}
 
In the calculation we use physical masses for the $D^{\prime(\ast)}$ mesons. In
Fig.~\ref{fig:ff} we show the different form factors evaluated in our model for the
$B\to D^{\prime(\ast)}$ transitions. For the sake of comparison we also show in a
different panel the corresponding ones for the $B\to D^{(\ast)}$ transitions. We see
that, even for the actual heavy quark masses the relations, in Eqs.~(\ref{eq:xi})
and~(\ref{eq:0}) are approximately satisfied for the $B\to D^{(\ast)}$ case over the
whole $w$ range. Deviations are expected due to the finite heavy quark masses and the
difference between $m_{b}$ and $m_{c}$. For the $B\to D^{\prime(\ast)}$ decays  those
differences are magnified by the fact that in the infinite heavy quark mass limit we
have $\xi(1)=0$ in this case.

\begin{figure}[!t]
\resizebox{7.cm}{!}{\includegraphics{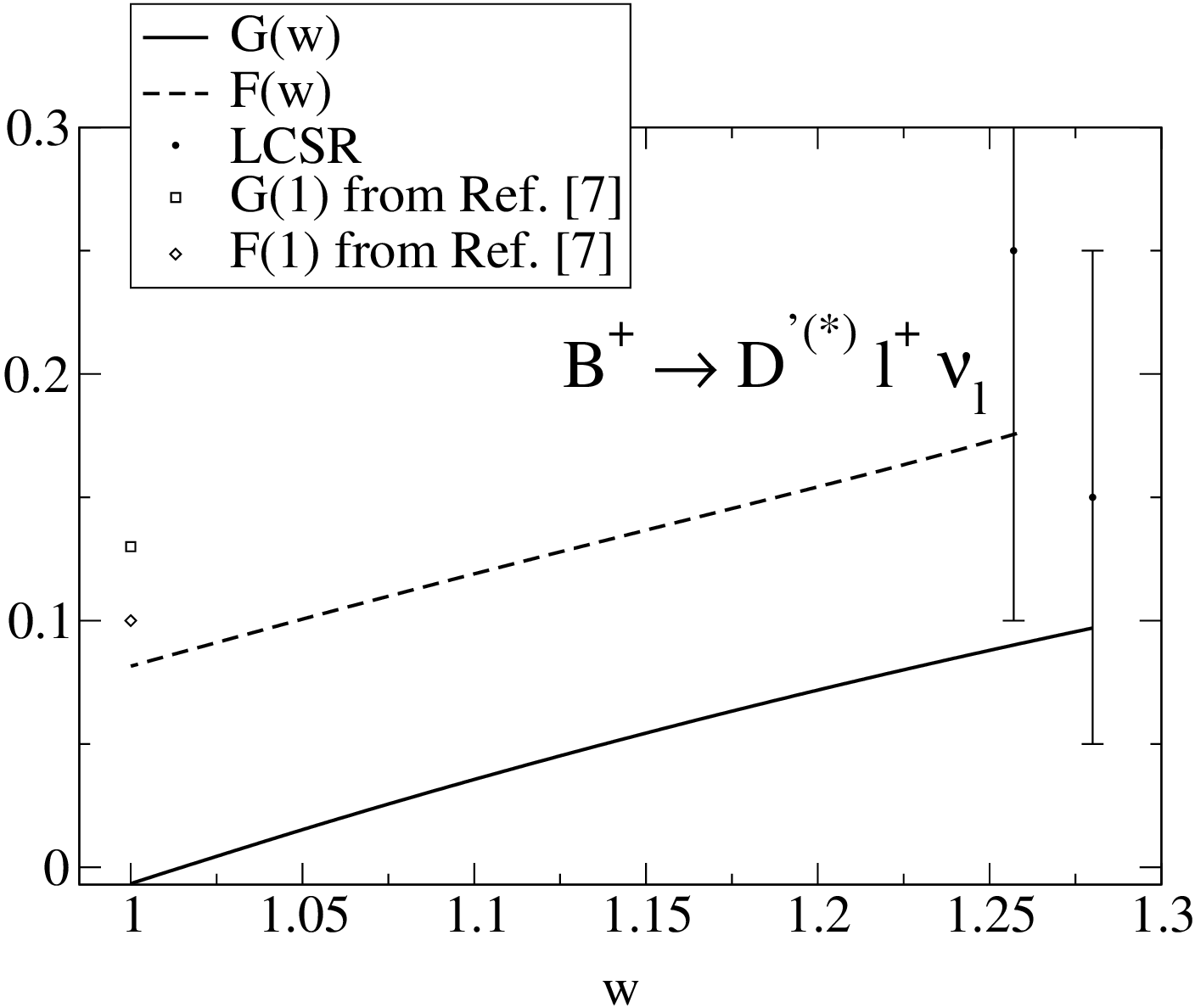}} 
\caption{\label{fig:fg} $G(w)$ and $F(w)$ factors for the  $B^{+}\to
D^{'}l^{+}\nu_{l}$ and $B^{+}\to D^{\prime\ast}l^{+}\nu_{l}$ transitions. We show
with error bars the estimates for maximum recoil obtained in
Ref.~\cite{Bernlochner:2012bc} adapting the light cone sum rule calculation of
Ref.~\cite{Faller:2008tr}. We also show the $1/m_Q$ corrected $G(1)$ and $F(1)$
values obtained in Ref.~\cite{Ebert:1999ed} using the relativistic quark model.}
\label{fig:GF}
\end{figure}

The $F(w)$ and $G(w)$ factors are depicted in Fig.~\ref{fig:fg}. Our results at
maximum recoil, $F(w_{\rm max})=0.2$ and $G(w_{\max})=0.1$, are compatible with the
estimates in Ref.~\cite{Bernlochner:2012bc}
\begin{equation}
F(w_{\rm max}) = 0.25\pm0.15, \quad \quad G(w_{\max}) = 0.15\pm0.1
\end{equation}
obtained adapting the light cone sum rule (LCSR) calculation of
Ref.~\cite{Faller:2008tr}. 

Integrating the differential decay width we obtain ${\cal B}(B^{+}\to
D'l^{+}\nu_{l})=(0.012\pm0.006)\%$ and ${\cal B}(B^{+}\to
D^{\prime\ast}l^{+}\nu_{l})=(0.097\pm0.015)\%$, where theoretical uncertainties have
been estimated varying the different model parameters within $10\%$ of their
central values. For the case of total strong decay widths calculated in
Tables~\ref{tab:D2550} and~\ref{tab:D2600}, we have seen that this $10\%$ variation
in the parameters induces changes which are smaller than the experimental error
bars. For the sum of the two semileptonic branching ratios, we thus obtain ${\cal
B}(B^+\to D^{\prime(\ast)}l^+\nu_l)=(0.109\pm0.016)\%$. Although this branching
ratio is larger than those found by Refs.~\cite{Suzuki:1993iq,Scora:1995ty} it is
still a factor of ten smaller than the expectation in Ref.~\cite{Bernlochner:2012bc}.
By looking at the individual ratios we also find that ${\cal B}(B^+\to D'l^+\nu_l)$
is smaller than ${\cal B}(B^+\to D^{\prime\ast}l^{+}\nu_{l})$, in  agreement with
Refs.~\cite{Suzuki:1993iq,Scora:1995ty}.

Our results for $G(w)$ and $F(w)$ at zero recoil differ from the values obtained in
Ref.~\cite{Ebert:1999ed} using the relativistic quark model. The discrepancy, clearly
visible in Fig.~\ref{fig:GF}, explains why our branching ratios are much smaller than
the ones in Ref.~\cite{Ebert:1999ed}. The change is larger for $F(1)$ which also
explains why ${\cal B}(B^{+}\to D'l^{+}\nu_{l})>{\cal B}(B^{+}\to
D^{\prime\ast}l^{+}\nu_{l})$ in Ref.~\cite{Ebert:1999ed}.

Concerning the non-leptonic decay our theoretical result, obtained within the
factorization approximation, is
\begin{equation}
\frac{{\cal B}(\bar B^{0}\to D^{\prime+}\pi^{-})}{{\cal B}(\bar B^{0}\to
D^{+}\pi^{-})} = 0.011\pm0.004,
\end{equation}
which combined with the experimental value for ${\cal B}(\bar B^{0}\to D^{+}\pi^{-}) =
(2.68\pm 0.13)\times10^{-3}$ gives for ${\cal B}(\bar B^{0}\to D^{\prime +}\pi^{-}) \sim
3\times 10^{-5}$, sufficiently large to be measured experimentally.

\section{CONCLUSIONS}
\label{sec:conclusions}

We have performed a theoretical calculation of the branching ratio for the $B^{+}\to
D^{\prime(\ast)}l^{+}\nu_{l}$ decays within the framework of the constituent quark
model described in Ref.~\cite{vijande2005constituent}. 

We find branching ratios much smaller than expectations in
Ref.~\cite{Bernlochner:2012bc}. However, the results are sensitive to the amount of
orthogonality between the $B$ and $D^{\prime(\ast)}$ wave functions and the latter
depend on the quark model used. In order to check that sensitivity, we have performed
an estimation of the theoretical uncertainties in our calculation, finding individual
branching ratios can change at most by $50\%$. Our results would not then confirm
that these contributions can explain the difference between the inclusive ${\cal
B}(B^{+}\to X_{c}l^{+}\nu)$ rate and the various exclusive channels.
 
Concerning the $1/2$ versus $3/2$ puzzle there is no need for a large branching ratio
into the $D^{\prime(\ast)}$ states to solve it. In fact, it was already shown in
Ref.~\cite{segovia2011semileptonic} that our model predicts similar $B$ semileptonic
decay rates into the two $j_{q}^{P}=1/2^{+}$ and $j_{q}^{P}=3/2^{+}$ doublets, being
our results for the different channels in agreement with experiment. To us, this
apparent puzzle appears only when one works in the infinite heavy quark mass limit
and neglects corrections on the inverse of the heavy quark masses.

We have also evaluated, in factorization approximation, the non-leptonic ${\cal
B}(\bar{B}^{0}\to D^{\prime+}\pi^{-})$ branching fraction. The latter reaction may
give additional information on the size of the form factors involved in the
semileptonic decay~\cite{Bernlochner:2012bc,Becirevic:2013mp} provided it can be
measured in $B$-factories or at LHCb in the near future.

\begin{acknowledgments}
This work has been partially funded by Ministerio de Ciencia y Tecnolog\'ia under
Contract Nos. FPA2010-21750-C02-02 and FIS2011-28853-C02-02, by the European
Community-Research Infrastructure Integrating Activity 'Study of Strongly
Interacting Matter' (HadronPhysics3 Grant No. 283286) by the Spanish
Ingenio-Consolider 2010 Program CPAN (CSD2007-00042) and by U.\,S.\ Department of
Energy, Office of Nuclear Physics, contract No.~DE-AC02-06CH11357.
\end{acknowledgments}


\bibliographystyle{apsrev}
\bibliography{weak_paper3}

\end{document}